\documentclass[11pt]{article}
\usepackage[a4paper, margin=1.2in]{geometry}
\usepackage{amsmath}
\usepackage{graphicx}
\usepackage{natbib}
\usepackage{amsthm,amssymb, color, booktabs}
\usepackage{url}

\def\bx{\boldsymbol{x}}
\def\by{\boldsymbol{y}}

\def\bH{\boldsymbol{H}}
\def\bI{\boldsymbol{I}}

\def\bQ{\boldsymbol{Q}}
\def\bR{\boldsymbol{R}}

\def\bU{\boldsymbol{U}}
\def\bV{\boldsymbol{V}}

\def\bX{\boldsymbol{X}}

\def\bSig{\boldsymbol{\Sigma}}
\def\bold1{\boldsymbol{1}}
\def\E{{\mathsf{E}}}

\def\cor{{\mathsf{corr}}}

\def\mR{{\mathbb{R}}}

\begin{document}


  \title{\bf Cross-Leverage Scores for Selecting\\Subsets of Explanatory Variables}
  \author{Katharina Parry,\\ 
    School of Fundamental Sciences, Massey University\\ ~ \\
    Leo N. Geppert, Alexander Munteanu, and Katja Ickstadt\thanks{
    	The authors gratefully acknowledge the funding by the Deutsche For\-schungs\-ge\-mein\-schaft (DFG) within the Collaborative Research Center SFB 876 ``Providing Information by Resource-Constrained Analysis", project C4, and by the Dortmund Data Science Center (DoDSc)}\\
    Department of Statistics, TU Dortmund University}
  \maketitle

\vspace*{.1in}

\begin{abstract}
In a standard regression problem, we have a set of explanatory variables whose effect on some response vector is modeled.  For wide binary data, such as genetic marker data, we often have two limitations. 
First, we have more parameters than observations. 
Second, main effects are not the main focus; instead the primary aim is to uncover interactions between the binary variables that effect the response. Methods such as logic regression are able to find combinations of the explanatory variables that capture higher-order relationships in the response. However, the number of explanatory variables these methods can handle is highly limited.
To address these two limitations we need to reduce the number of variables prior to computationally demanding analyses. In this paper, we demonstrate the usefulness of using so-called cross-leverage scores as a means of sampling subsets of explanatory variables while retaining the valuable interactions.\\

\end{abstract}

\vspace*{.3in}

\section{Introduction}

We present a new approach to reducing the number of variables in very wide binary data sets based on a sampling approach using cross-leverage scores. There are two characteristics of wide binary data that often limit the statistical anal\-y\-ses available. First, the number of observations is considerably lower than the number of parameters. Second, the variables have all been measured on a binary scale. A typical example is given by genetic marker data, where we tend to have a small number of individuals for which thousands of binary variables have been measured. 
An adaptation of standard regression models, called logic regression, was devised by \citet{Ruczinski2003} as a means to determine the relationship between binary explanatory variables and a response of interest. 
This new method has proven successful at providing useful models in many areas such as medical informatics and economics, e.g. \citet{Rathod2015,Vesely2016}. 

One limitation, however, is the number of variables this method can currently handle. The reason for this lies in the fact that the number of possible combinations of the main effects as well as their interactions is very large, especially in relation to the number of observations \citep{Schwender2007,Tietz2019}.

A consequence of this drawback is that we need to reduce the number of variables prior to the analysis. The challenge is two-fold, on the one hand little information should be lost due to the reduction to produce approximately accurate estimates, and on the other hand, the reduction step needs to be efficient enough to work well for large data sets. 

This has led to the development of sophisticated sampling algorithms, where the original data matrix $\bX$ is reduced by selecting rows or columns. Uniform sampling is known to work well if the data matrix can be assumed to have low coherence \citep{Kumar2012, Cohen2014}. In other words, the singular vectors of the data matrix are not essentially correlated with the vectors that define its standard basis \citep{Mohri2011}. In this case, the norm of the matrix is spread evenly across all dimensions. However, this assumption is very limiting and breaks, e.g., for sparse data, cf. \citet{CandesR2007}. 

In their seminal work, \citet{Drineas2006} introduce an importance sampling approach for least-squares regression. Their approach is based on the row-norms of a matrix $\bU$ which is derived from the singular value decomposition of the data matrix $\bX=\bU\bSig\bV^T$. These row-norms coincide with what is known as \emph{leverage scores} dating back to \citet{Cook77}. 

Leverage scores are well known in statistics as a measure of how unusual an observation is, and continue to enjoy enduring popularity in many applications, e.g. outlier detection \citep{Chatterjee1986,Gruttola1987a,Rousseeuw1990,Rousseeuw2011}. In contrast, leverage scores have become widely used in computer sciences as a tool in importance sampling methods \citep{Drineas2006,Drineas2008,Drineas2010,Li2013}. In both areas, the idea is that a higher leverage score indicates that row $\bx_i$ is more important in composing the spectrum of $\bX \in \mR^{n\times p}$. Consequently, for $p<n$, the reduced matrix is obtained by sampling $O(p\ln p)$ rows or columns with probabilities proportional to their leverage scores.

While leverage scores have been found to be immensely useful, there remained the issue that reduction by decomposing a matrix is still computationally intensive. In fact, reducing the matrix in a na{\"i}ve way before fitting the regression model is as slow as solving the original least-squares regression problem! This was posed as an open problem in \citet{Drineas2006} and consequently computer science research has improved the efficiency of these algorithms in multiple ways \citep{Halko2011,Drineas2011,Clarkson2013,Nelson2013}. Note that if the subsequent analyses are computationally more demanding than least-squares fitting, like e.g. Markov Chain Monte Carlo sampling or logic regression, then the computational effort for computing probabilities and sampling as a pre-processing step becomes negligible \citep{Geppert2017}.

Despite these recent advances in improving computational efficiency when using leverage scores, \emph{cross-leverage scores} \citep{ChatterjeeHadi1988} have not yet been studied for variable selection and in particular for uncovering \emph{interaction} terms, i.e., combinations of variables that together form a relationship with the response.\\

The overall aim of this paper is thus to introduce cross-leverage scores as a new variable selection measure for uncovering interactions. 
Specifically, our contributions can be summarized as follows:

\begin{enumerate}
	\item We study the properties of both leverages and cross-leverages for variable selection on binary data and show differences between the two in how they express themselves.
	\item We show that cross-leverages supersede leverages, as well as other common variable selection measures, in terms of finding meaningful interactions.
	\item We formulate an improved variable selection measure which involves combining leverage scores with cross-leverage scores.
	\item We study the pre-selection of variables when applying logic regression and evaluate this method on simulated data as well as data from the HapMap project\footnote{The original web site www.hapmap.org is unfortunately offline, instead, the following site contains information on the HapMap project and data resources: https://www.genome.gov/11511175/about-the-international-hapmap-project-fact-sheet}.
\end{enumerate}

Our paper is structured as follows. In the next section we provide details on lever\-age/cross-leverage scores, and then define an importance sampler which is based on these measures as well as a number of other commonly used selection criteria. Our methods section then comes to an end with a brief introduction to logic regression. In section 3, we describe how we generate data and the set-up of our simulation study, as well as the real-world data set we work with to illustrate our method. In section 4 we present our results, and then conclude with some final comments and ideas for future work in section 5.

\section{Methods}
First, we explain what leverage and cross-leverage scores are, and how they can be efficiently computed. We then describe importance sampling approaches where we consider four different ways of selecting variables. A brief introduction to logic regression concludes this section.

\subsection{Leverage and cross-leverage scores}

Leverage scores are usually defined on the basis of observations. As we consider the case of wide data, where we have more parameters $p$ than observations $n$, we need the data matrix to be transposed. This switches the roles of observations and variables. Deviating from usual regression modeling, the matrix from which we derive the scores is a combination of the data matrix $\bX$ and the response vector $\by$, given as 
\begin{equation*}
\tilde{\bX}=[\bX, \by]^T \in \mR^{{(p+1)} \times n}.
\end{equation*}


Now, the matrix from which we can derive our scores is 
\begin{equation*}
\bH =\tilde{\bX} (\tilde{\bX}^T\tilde{\bX})^{-1}\tilde{\bX}^T \in \mR^{(p+1)\times (p+1)},
\end{equation*}
which is a projection matrix for the column space of $\tilde{\bX}$ that gives us information about the impact of the variables (instead of observations) on the projection and is commonly referred to as \emph{hat matrix}, cf. \citet{HoaglinW1978}. Please note that interpretations from standard least squares problems cannot easily be transferred to our case.

Calculating this matrix algebraically involves computing the inverse of $\tilde{\bX}^T\tilde{\bX}$, which can lead to numerical instabilities, as discussed in \citet{Geppert2017}. Therefore, following the ideas presented in \citet{Drineas2011} we use a QR-decomposition \citep{GolubVanLoan2013} to obtain the values in a numerically stable way. 

The QR-decomposition, $\tilde{\bX} = \bQ\bR$, yields an orthonormal matrix $\bQ\in \mathbb{R}^{(p+1)\times n}$, such that $\bQ^T\bQ = \bI$ and an invertible upper triangular matrix $\bR\in\mathbb{R}^{{n}\times {n}}$ from which the hat matrix can be obtained as
\begin{align*}
\bH &= \tilde{\bX} (\tilde{\bX}^T\tilde{\bX})^{-1}\tilde{\bX}^T \\
&= \bQ\bR (\bR^T {\bQ^T\bQ} \bR)^{-1} \bR^T\bQ^T \\
&= \bQ\bR \bR^{-1} (\bR^T)^{-1} \bR^T\bQ^T = \bQ \bQ^T.
\label{hat2} 
\end{align*}
If we calculate the hat matrix as defined above, it will have dimensions $(p+1) \times (p+1)$, where the final row and column relate to the response variable vector.

The first $p$ elements on its diagonal, $l_i=h_{ii}$ for $i\in\{1,\ldots,p\}$, provide information on the importance of the $i^{th}$ variable, $\bx_i$, in composing the spectrum of $\tilde{\bX}$, cf. \citet{Drineas2011}. These measures are called \emph{leverage scores} (LS) and coincide with the squared row-norms $l_i=\|\bQ_{i\cdot}\|^2_2$ of the orthonormal matrix.

Another measure called \emph{cross-leverage scores} (CLS) is given by the off-diagonal elements of the hat matrix \citep{ChatterjeeHadi1988}. These values, denoted as $c_{ij}=h_{ij}, i \neq j$, provide information on the mutual influence of the $i^{th}$ variable, $\bx_i$, and the $j^{th}$ variable, $\bx_j$. Most of the focus in this work will lie on the CLS of the form $c_{ij}$, $i\in\{1,\ldots,p\}$, $j=p+1$, measuring the influence of the $i^{th}$ variable, $\bx_i$ on the response $\by$.

The computational cost of computing $\bH$ as described above in the case $p\geq n$ is $O(pn^2)$. The numerically more stable QR-decomposition is no remedy to this. Indeed, computing the LS or CLS in $o(pn^2)$ time was posed as an open question in \citet{Drineas2006} and was not resolved before \citet{Drineas2011}, who were the first to approximate least-squares regression based on sub-sampling via LS faster than the time needed to solve the problem exactly via standard methods. Due to \citet{Drineas2011} the LS and the large CLS in absolute value can be accurately approximated in $O(pn\ln p)$ time. More recent advances in the random projection regime have further improved this to $O({\mathbf{nnz}}(\bX)+\operatorname{poly}(n))$ \citep{Clarkson2013,Nelson2013}, where ${\mathbf{nnz}}(\bX)$ is the number of non-zeros in the data, referred to as \emph{input-sparsity}, and is bounded by $O(np)$ even in the case of dense data. Those methods also support extensions to massive data models where the data matrix can be presented in arbitrarily ordered additive updates to its entries, see \citet{Geppert2017} for a comprehensive discussion on streaming and distributed massive data.

\subsection{Importance sampling}
In our importance sampler, the aim is to select a subset of variables that include as many relevant ones as possible. There are two factors we need to take into account. First, the number of variables to select, that is, the sample size. Second, how to define the selection weights.

Regarding the sample size, obviously the more original variables we include the more likely it is that we select all the influential variables. However, for the sake of efficiency we need to keep our sample size as small as possible. 
We take samples of size $O(n\ln n)$ following the guidance of \citet{BoutsidisMD09}, who show that using this number of selected variables retains the spectrum up to multiplicative errors and thus retains the full rank $n$ of the original matrix. 
We note that $\Omega(n \ln n)$ is also a lower bound \citep{Tropp11}, which indicates that we cannot do much better without imposing restrictive assumptions on the data, like low rank or low coherence.

With respect to the selection criteria, we consider four different schema in terms of extracting useful variables:
\begin{enumerate}
	\item Cross-leverage scores (CLS)
	\item Leverage scores (LS)
	\item Correlation (COR)
	\item P-value.
\end{enumerate}
The first two involve drawing samples using the raw leverage and cross-leverage scores. Explanatory variables with more extreme leverage/cross-leverage scores are selected.

The third option is to calculate the Pearson correlation as a measure of the dependency between each of the explanatory variables and the response. Those explanatory variables with the highest correlations between the original explanatory variable and the response are selected. This approach is used in feature pre-processing, e.g. \citet{Kohavi1997}. 

The fourth variable selection criterion involves fitting $p$ linear regression models for a single explanatory variable at a time and selecting those variables which show the most significance. That is, our final criterion involves selecting variables with the lowest p-value. Works such as \citet{Sun1996} have pointed out the limitations of this approach, but it continues to be widely used, hence its inclusion as a benchmark in our analysis. 

\subsection{Correlation versus cross-leverage scores}

Both cross-leverage scores and correlations are derived from the off-diagonals of similar, though not equal, matrices that describe the shape of a data set. This gives rise to the idea that these two types of summaries are measuring essentially the same thing. We present a counter-example that shows this is not the case. In the following example, cross-leverages are able to detect useful variables where correlations fail. 

We consider a toy example data set, shown in Table \ref{CLSvsCOR}. It has 8 observations and 30 variables of which the first four, $\bx_1,\ldots,\bx_4$, have an influence on the dependent variable $\by$. The response $\by$ takes values of 1 if $\bx_1=\bx_2=1$ or $\bx_3=\bx_4=1$. However, the data set is constructed in a way that ensures there is no correlation between either of the first two variables and the response, that is, $\cor(\bx_1, \by) = \cor(\bx_2, \by) = 0$. The third and fourth variable show moderate correlation of $0.26$ with the response variable.

\begin{table}[ht]
	\centering       
	\begin{tabular}{cccccccccccccccccccr}
		\toprule
		&\multicolumn{18}{c}{ Explanatory variable $\bx$} & \\
		1 & 2 & 3 & 4 & 5 & 6 & 7 & 8 & 9 & 10 & 11 & 12 & 13 & 14 & 15 & 16 & 17 &  ... & 30 & $\by$ \\ 
		\midrule
		1 & 1 & 0 & 1 & 1 & 0 & 1 & 0 & 1 & 0 & 0 & 0 & 0 & 1 & 1 & 0 & 1 &...  & 1 & 1 \\ 
		1 & 1 & 1 & 0 & 0 & 0 & 0 & 0 & 1 & 1 & 1 & 0 & 0 & 1 & 0 & 1 & 1 &...  & 0 & 1 \\ 
		0 & 0 & 1 & 1 & 1 & 1 & 0 & 1 & 0 & 0 & 1 & 0 & 1 & 1 & 1 & 1 & 1 &...  & 1 & 1 \\ 
		0 & 0 & 1 & 1 & 0 & 1 & 1 & 1 & 0 & 0 & 1 & 0 & 0 & 1 & 0 & 0 & 1 &...  & 0 & 1 \\ 
		1 & 0 & 0 & 1 & 1 & 1 & 1 & 1 & 1 & 1 & 1 & 0 & 1 & 0 & 0 & 1 & 1 &...  & 0 & 0 \\ 
		1 & 0 & 1 & 0 & 0 & 0 & 1 & 0 & 1 & 0 & 1 & 0 & 0 & 1 & 0 & 0 & 0 &...  & 0 & 0 \\ 
		0 & 1 & 0 & 1 & 1 & 0 & 1 & 0 & 1 & 1 & 1 & 0 & 0 & 1 & 1 & 1 & 1 &...  & 0 & 0 \\ 
		0 & 1 & 1 & 0 & 1 & 0 & 1 & 0 & 1 & 0 & 0 & 0 & 0 & 0 & 1 & 0 & 0 &...  & 0 & 0 \\ 
		\bottomrule
	\end{tabular}
	\caption{Toy example data where strong relationships exist between the first four variables, albeit the correlation between the first two with the response is zero. \label{CLSvsCOR}}
\end{table}

Because $n=8$, $\lceil 8\ln(8)\rceil = 17$ variables are chosen according to the respective correlation or cross-leverage score values. In this toy example, variable selection via correlation chooses the third and fourth variable, but the first two are systematically not found as the correlation always equals 0. In contrast, for the given data set, all four important variables are selected using their cross-leverage scores.

This toy example demonstrates that while pre-selecting variables according to their cross-leverage scores and their correlation with the dependent variable can lead to similar results, there are cases in which selection according to the correlation systematically does not and cannot find the important variables, and is thus conceptionally different.

\subsection{Logic regression analysis}
This method is adapted from the standard generalized linear model regression approach to handle binary data. The aim of these models is to explicitly identify interactions among the explanatory variables. That is, instead of using the binary predictors as given, this method combines them using logic operators $\land $ (and) and $\lor$ (or). The ! operator denotes the negation of a predictor, e.g. if $\mathcal E$ denotes the occurrence
of an event, then $!\mathcal E$ denote its absence. These resulting Boolean expressions form so-called logic trees, where the explanatory variables that make up each tree are called leaves. In general a model fitted in this manner is as follows
\begin{equation}
g(\E(\by))= \beta_0 + \sum_{i=1}^{T} \beta_i L_i,
\label{logicglm}
\end{equation}
where $g()$ is the logit link and $L_i$, for $i\in\{1,\ldots,T\}$ are the Boolean combinations of the original binary predictors that explain the variation in the response. To avoid ambiguity, note that every Boolean formula $L_i$ can be expressed in a canonical form called disjunctive normal form \citep{Whitesitt2012}, which allows for interpretability and is particularly suitable to be represented as a logic tree.

If the response variable is binary as well then we are doing a classification where there is only one tree in the model. In this case, we can simplify Equation \ref{logicglm} to a model that can be described as 
\begin{equation}
g(\E(\by))= \beta_0 +  \beta_1 L_1.
\label{onetree}
\end{equation}

Fitting this model, as first described in \citet{Ruczinski2003}, involves using an iterative search algorithm based on Simulated Annealing. The algorithm starts with a single binary explanatory variable (or leaf). Each move is defined as a modification of the tree(s) in the model based on six standard operations such as adding or replacing a variable (leaf), as well as the switching of Boolean operators. The considerable number of combinations due to each of these operations that are possible give rise to an extremely large number of possible models for a given set of $p$ explanatory variables. In particular, the number of possible models can be shown to increase as a doubly exponential function of $p$.

A method developed by \citet{Schwender2007} called LogicFS is also iterative and builds on the above work. It is an ensemble method which involves bootstrapping the data a suitably large number of times and at each iteration fitting a logic regression model to the in-bag bootstrap training set to obtain stability. LogicFS includes the frequency of occurrence of included explanatory variables. These are meant to assist in identifying useful predictors. Another recent work \citep{Tietz2019} introduces SurvivalFS which extends the previous method to binary data associated with survival times.


\section{Data description}
The first part of this section is a breakdown of our R function used to create artificial data for our simulation study in R, version 3.6.0 \cite{RCore360}. The code will be available from the authors upon request. This data allows us to demonstrate the extent to which leverage/cross-leverage scores are able to uncover variables that are particularly relevant for interactions as well as main effects in the model. In the second part of this section, we describe our real-world data set.

\subsection{Simulated data}
We create artificial data where we know the underlying model that generated the measurements. These models may consist only of main effects, as well as the more realistic scenario where particular combinations or interactions of variables lead to a disease occurring in the individual.

The form of the simulated data sets are matrices with $n$ rows and $p+1$ columns. The rows correspond to $n$ observations, and the columns correspond to $p$ explanatory variables and the response variable column. All explanatory variables as well as the response variable are binary. 

In practice, binary data can occur in multiple settings. One example is data that contains information on sites in the human DNA sequence where individuals differ at a single DNA base. These sites are called single-nucleotide polymorphisms (SNPs). This kind of data typically has less observations than variables, that is $n < p$. The response is a measure of the disease status of the $i^{th}$ individual, which is defined as $1$ if the individual is diseased, and $0$ otherwise. The explanatory variables detail for all individuals a $1$ if a particular genetic marker is present or $0$ otherwise. Other examples are variables that indicate whether the concentration of a substance could be measured or whether the measurement was below the limit of detection as well as measurements on a Likert scale with an even number of items, that are consequently dichotomized. 

Our R function for generating data has four arguments, {\tt n}, {\tt p}, {\tt vars}, and {\tt probs}. The number of observations {\tt n}, the number of explanatory variables {\tt p}, in which the response variable is not included, the list of relevant explanatory variables and interactions thereof {\tt vars}, and the probabilities associated with the explanatory variables {\tt probs}.

By default the explanatory variables are assumed to follow a Bernoulli distribution with parameter 0.5, that is, if they have no effect on the response then the process of having a value of 1 for any given explanatory variable is akin to flipping a coin. 
The aim of the simulated data is to produce a comparable number of disease cases to controls. Obviously, if all variables were Bernoulli(0.5) distributed then it would become very unlikely that we observe a 1 if for example, a set of four variables all need to simultaneously be 1 in order for this to occur. The argument {\tt probs} allows us to adjust the probabilities of certain explanatory variables linked to getting a disease status of 1 such that the number of cases, $n_1$ match the number of controls, $n_2$, where $n_1 + n_2 = n$.

Our first set of simulated data was generated in order to investigate the distributions of leverage/cross-leverage scores as an initial exploration of how they might be used for variable selection. We are particularly interested in the extent to which there is a distinction in how the scores express themselves if there are relationships between the variables.

In all these data sets, we have 60 observations, and 1\,000 explanatory variables of which the first 10 have an effect on the response. Note that defining the first 10 variables as influential has no impact on the selection step as there is no inherent order in categorical data. We consider three scenarios where $y=1$ if:
\begin{description}
	\item[Scenario 1] One of first 10 variables is equal to 1, that is, $L_1 = S_1 \lor  S_2 \lor \cdots \lor S_{10}$.
	\item[Scenario 2] One of the five following conditions hold:\\ 
	$L_2 = (S_1 \land S_2 \land S_3) \lor (S_4 \land S_5 \land S_6) \lor (S_7 \land S_8) \lor S_9 \lor S_{10}$.
	\item[Scenario 3] One of the four following conditions hold:\\ 
	$L_3 = (S_1 \land S_2 \land S_3 \land S_4) \lor (S_5 \land S_6\land S_7) \lor (S_8 \land S_9) \lor S_{10}$.
\end{description}

The contrast between Scenario 1 and the other two allows us to look at the possible differences in how we should approach variable selection for data where we have only main effects, and data where we have varying degrees of higher-order interaction terms. 

A second set of data was simulated with the intent of exploring the effect on lever\-age/cross-leverage scores when increasing either the number of observations or variables. We allowed the sample size to increase from $n=60$ to $n= 100$, $n=300$ and then $n= 600$. For the variables, we first double their number from $p=1\,000$ up to $p=2\,000$, then consider a six-fold increase to $p=6\,000$. Table \ref{table1} gives an overview of the two sets of simulated data.

\begin{table}[h]
	\centering
	\normalsize
	\begin{tabular}{lcccl}
		\toprule 
		\rule[-1ex]{0pt}{2.5ex}  Set & $n$ & $p$ & Scenario & Interactions \\ 
		\midrule 
		\rule[-1ex]{0pt}{2.5ex} 1 & 60 & 1\,000 & 1 & Main effects only   \\ 
		\rule[-1ex]{0pt}{2.5ex} 1 & 60 & 1\,000  & 2 & Up to 3-way  \\ 
		\rule[-1ex]{0pt}{2.5ex} 1 & 60 & 1\,000 & 3 & Up to 4-way   \\ 
		\rule[-1ex]{0pt}{2.5ex} 2 & 60 & 2\,000  & 3 & Up to 4-way  \\ 
		\rule[-1ex]{0pt}{2.5ex} 2 & 60 & 6\,000  & 3 & Up to 4-way \\ 
		\rule[-1ex]{0pt}{2.5ex} 2 & 100 & 1\,000  & 3 &  Up to 4-way  \\ 
		\rule[-1ex]{0pt}{2.5ex} 2 & 300 & 1\,000  & 3 & Up to 4-way  \\ 
		\rule[-1ex]{0pt}{2.5ex} 2 & 600 & 1\,000 & 3 & Up to 4-way  \\ 
		\bottomrule
	\end{tabular} 
	\vspace{0.1cm}
	\caption{Overview of various settings considered when simulating data. The first set focuses on differences between how the LS and CLS of influential variables express themselves. The second set considers the effects of increases in $n$ and $p$ on the scores.}\label{table1}
\end{table}

In all cases, we simulate 10\,000 data sets for the given settings and calculate their scores. We then plot their distributions as kernel densities. 

\subsection{HapMap data}
\label{sec:HapMap}
We consider genetic data that was collected as part of an international collaboration with the aim of developing a haplotype map of the human genome, hence its name HapMap data. We consider a subset that is available as part of the R package ``SNPassoc'' \citep{Gonzalez2014}. This data set is a list of haplotype blocks and the genetic variations, called single-nucleotide polymorphisms (SNPs) that identify them, from various populations in different regions around the world. We chose to work with this data as it represents a typical real-world example of the data our variable selection approach was developed for. That is, the data is binary, thus lending itself to model fitting using logic regression, and contains considerably more variables than observations. 

That is, we have $p=9307$ SNPs, our explanatory variables, for $n=120$ individuals from two separate ethnic groups. In this case the response variable is 1 if an individual is from Europe, and 0 if they are one of the Yorùbá people, an ethnic group that inhabits western Africa.

We make some comments on the data pre-processing: We use the function {\tt additive} from the ``SNPassoc''  R package to re-encode the categorical entries in numeric form, where the values can be 0, 1 or 2. These three outcomes correspond to the different ways a genetic variation can occur \citep{Schwender2007}.

Around 4.39\% of the variables had missing values. We imputed these values using the marginal distributions of the given SNP. That is, the probability of an imputed value is matched to the relative frequency with which each of the possible outcomes occur.

Of the 9307 SNPs, 1657 did not express themselves for any of the individuals, that is, a zero was recorded in all cases. These SNPs were removed as they provide no information on differences in the response.

\section{Results}
We first present results from our simulation study, then contrast our variable selection approach using cross-leverage scores to existing methods for a real-world data set.  

\subsection{Effect of variable relationships on leverage/cross-leverage scores}

\begin{figure}[h!] 
	\includegraphics[width=\textwidth]{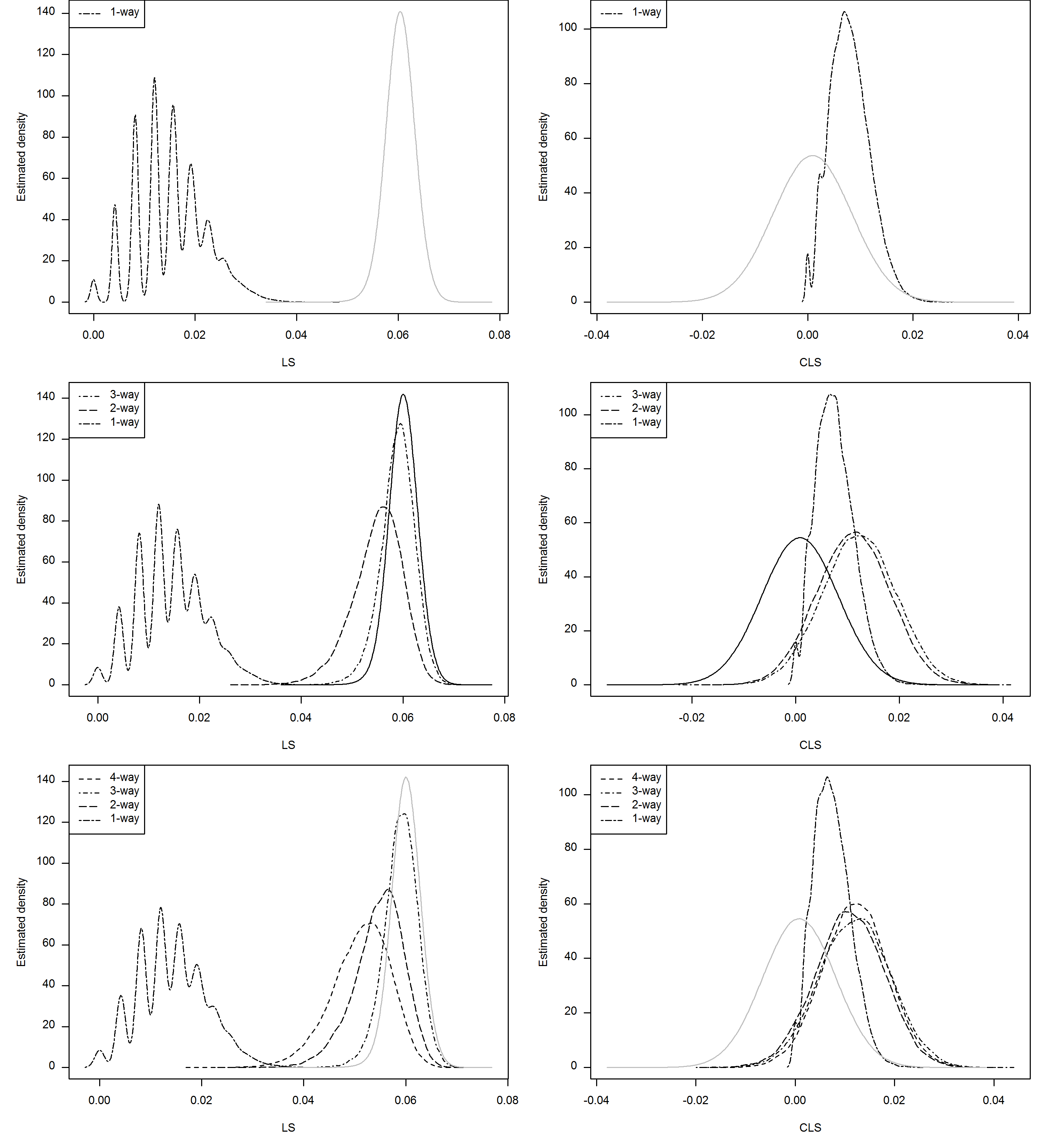}\\
	\caption[]{Kernel densities of (C)LS from 10\,000 simulated data sets. Top, middle and bottom panel relate to Scenarios 1 to 3, respectively. Plots in left column depict leverage scores (LS), and right-hand plots depict cross-leverage scores (CLS). The gray kernel density represents the collective distribution of all irrelevant variables.}
	\label{part1}
\end{figure}

In Figure \ref{part1} we show the distributions of leverage scores (left) and cross-leverage scores (right) for the data simulated under three different scenarios. 

One first impression is that the distribution of leverage scores for main effects is very different from the distribution of all other variables, irrelevant ones included. While the separation between leverages for main effects from the other scores is very clear, this comes at a cost. If higher-order effects, up to four-way, are present, then the leverage scores of the variables involved in these relationships are similar to the point of being confused with the scores of the irrelevant variables.  

Another striking characteristic is that leverage scores behave in a different manner for binary data than what is commonly seen for non-binary data. That is, while in the standard case of non-binary explanatory variables, a higher leverage score is associated with a useful variable, we see here the main effects have leverage scores that are lower than those of the other variables. When combining binary and continuous variables as influential independent variables, their resulting leverage scores are lower for the binary variables and higher for the continuous variables, compared to the non-relevant variables.

When viewing the right column we see that the cross-leverage scores have a lot more overlap between the distributions of relevant and irrelevant variables. However, in light of our primary interest in interactions, the considerably clearer separation of scores for main as well as higher-order effects from those of irrelevant variables can be considered advantageous.  

\subsection{Effect of $n$ and $p$}
Given that our main interest is to uncover relationships between variables, we will hence consider the scenario with up to fourth-order interactions. While holding the sample size fixed at $n=60$, we investigate the impact of a tenfold increase in the number of variables, see Figure \ref{part2a}:
\begin{figure}[h!] 
	\includegraphics[width=\textwidth]{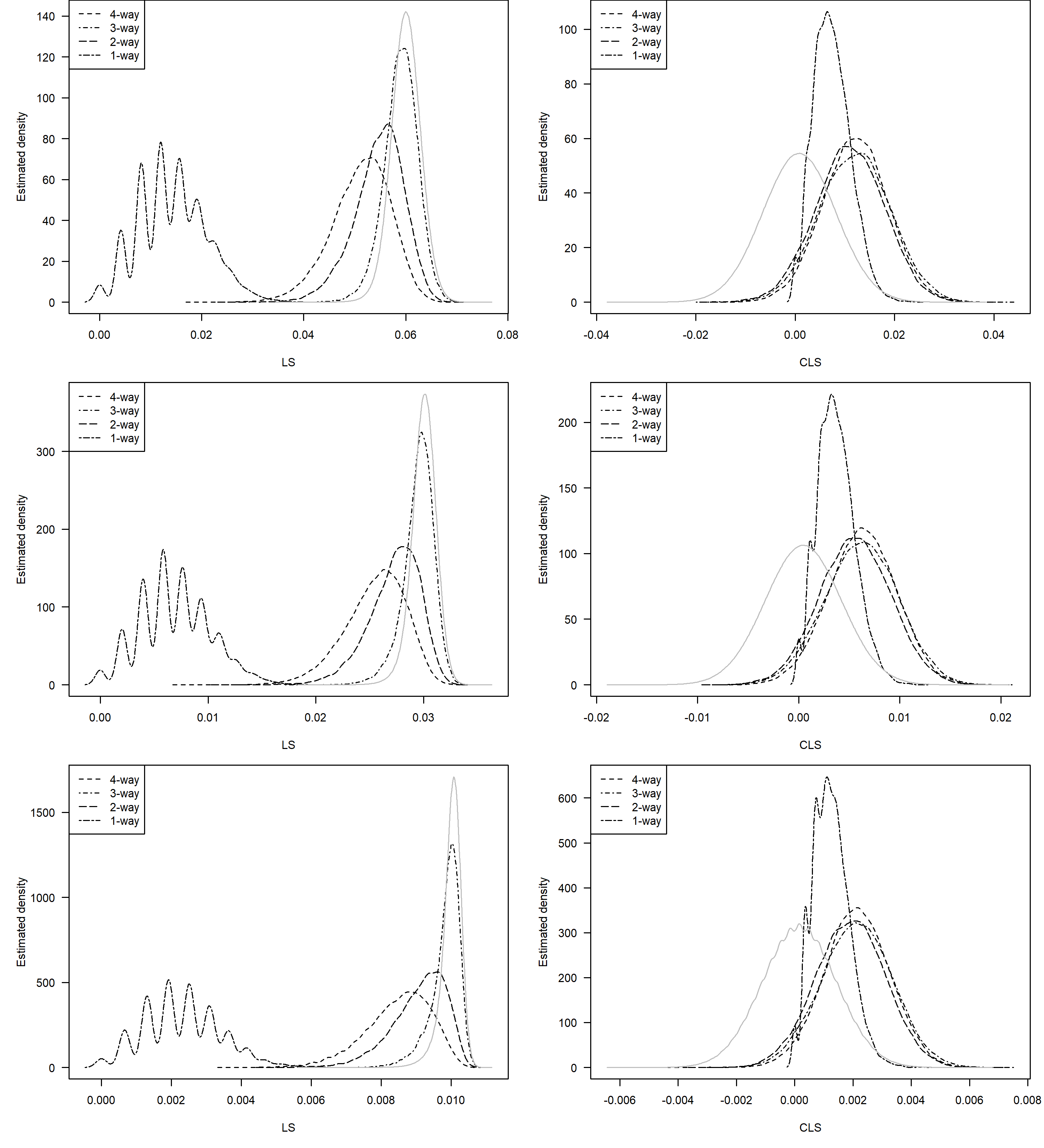}\\
	\caption[]{Effect of increasing the number of variables: Kernel densities of CLS from 10\,000 simulated data sets based on scenario 3. The number of observations was fixed at n=60, and the top, middle and bottom panel show results where p=1\,000, p=2\,000 and p=6\,000, respectively. Plots in left column depict leverage scores (LS), and right-hand plots depict cross-leverage scores (CLS). The gray kernel density represents the collective distribution of all irrelevant variables.}
	\label{part2a}
\end{figure}

Note that the tenfold increase relates to the number of variables considered, the number of relevant predictor variables remains the same at 10. The sample space containing a larger number of irrelevant variables mirrors real-life data of this nature more authentically. The increases lead to no improvement, which is to be expected given that the additional variables contain no useful information. Instead, we have introduced more noise. It is however promising that the distinctions between the interaction and main effect terms from irrelevant variables is overall not affected.

Now, let us consider the effect of increasing the number of observations.
\begin{figure}[h!] 
	\centering
	\includegraphics[width=0.87\textwidth]{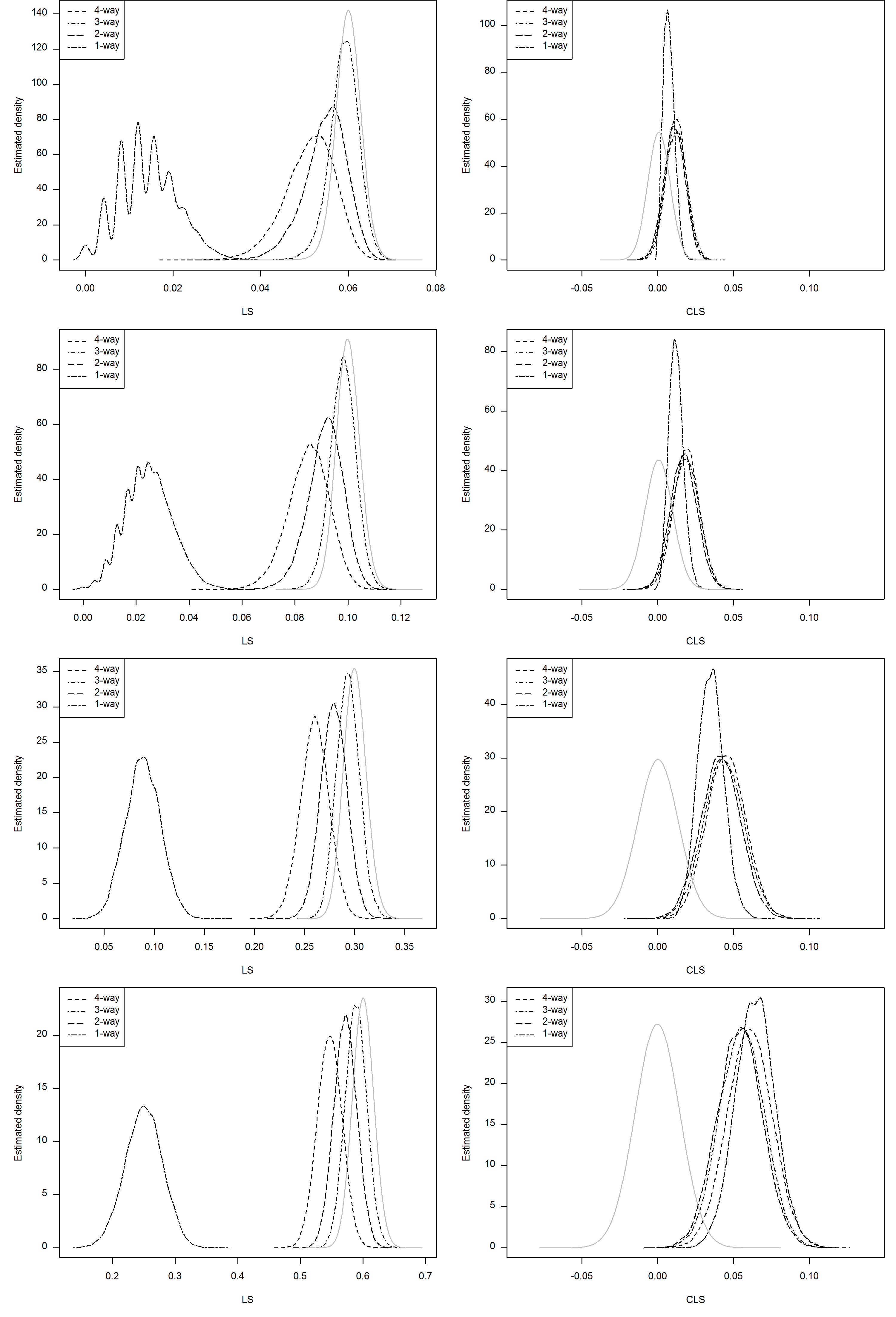}\\
	\caption[]{Effect of increasing the number of observations: Kernel densities of CLS from 10\,000 simulated data sets based on scenario 3. The number of variables was fixed at p=1\,000, and the plots from top to bottom correspond to n=60, n=100, n=300 and n=600, respectively. Plots in left column depict leverage scores (LS), and right-hand plots depict cross-leverage scores (CLS). The gray kernel density represents the collective distribution of all irrelevant variables.}
	\label{part2b}
\end{figure}
We observe in Figure \ref{part2b} a trend where as $n$ increases the overlap between  the average cross-leverage scores of higher-order terms as well as main effects and the average scores of the irrelevant terms decreases. This is good news for the variable selection process as this means that the scores for variables that are useful for prediction lie increasingly further away from the scores we can ignore. For example, as seen in the bottom panel of Figure \ref{part2b}, we can conclude that keeping only variables with a CLS greater than 0.025 would effectively remove all irrelevant variables.

In contrast, we see that the leverages of variables involved in higher-order interactions have been separated out as well, albeit to a far lesser extent than in the cross-leverage case. That is, while the leverage scores of irrelevant variables do space out a little more from important ones, there is still a considerable overlap between the distribution of the irrelevant variables with those involved in interactions.  

In summary, our initial exploratory analysis shows that cross-leverage scores distinguish variables of higher-order interactions better, and improve in their usefulness for variable selection more than leverage scores when increasing the sample size.

\subsection{Sampling success rate}
\label{res:samplingsuccess}

Here we compare the four selection criteria in terms of their ability to capture useful variables. To this end, we revisit the sets of data generated for the three scenarios, where we have $n=60$ and $p=1\,000$ of which 10 are useful explanatory variables. We take samples of size $\lceil n\ln(n) \rceil=246$, using the four different selection criteria. For each of 10\,000 data sets, we count for each criterion in each scenario the number of useful variables captured. 

\begin{figure}[h!] 
	\includegraphics[width=\textwidth]{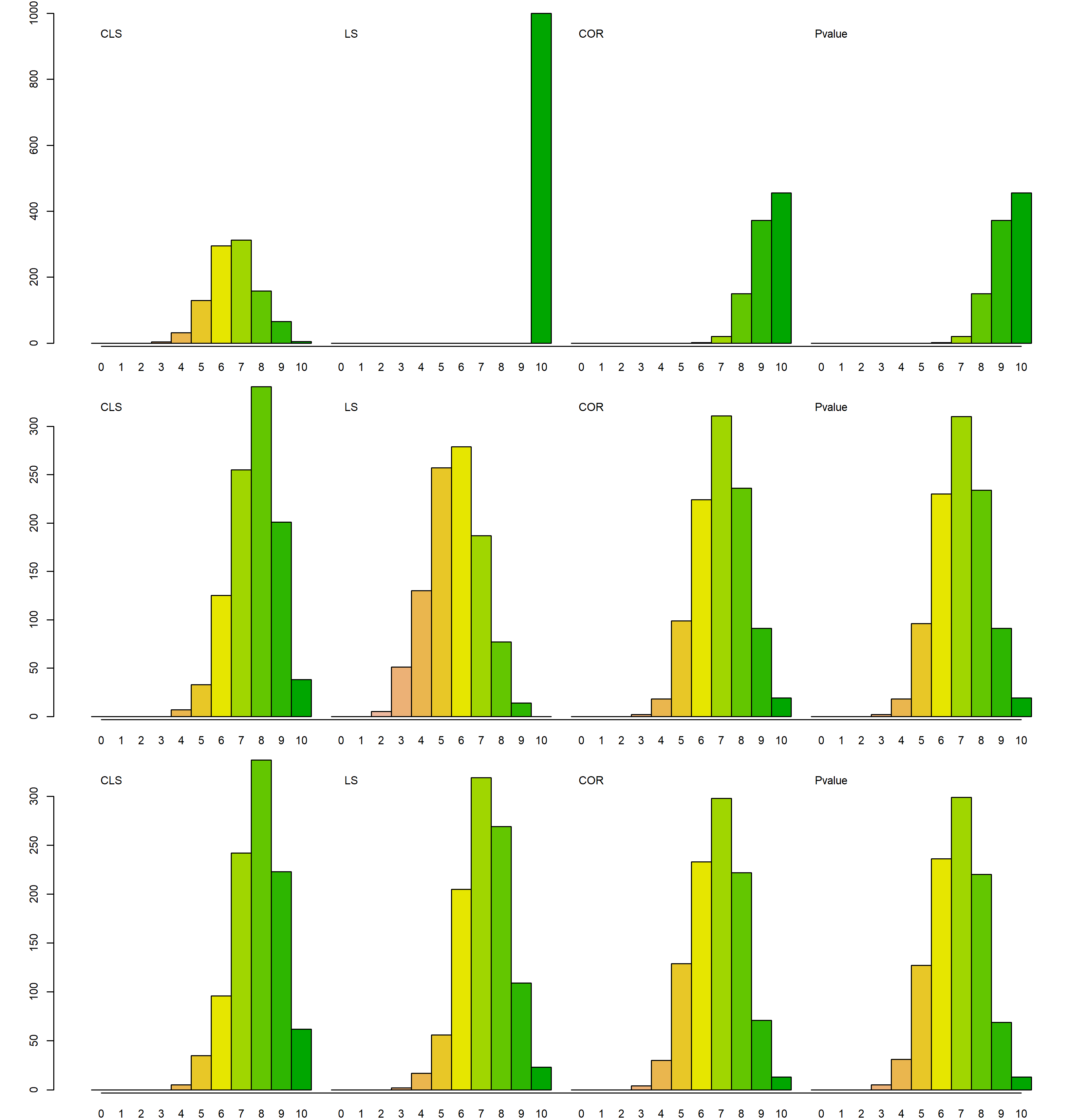}\\
	\caption[]{Histograms showing number of useful variables selected using each criterion, CLS, LS, COR and p-value (from left to right). The success of each selection criterion is shown for scenarios 1 to 3 in the top, middle and bottom row, respectively. The scale on the x axis runs from 0 to 10, where 0 corresponds to a subset that contains no useful explanatory variable, and 10 to a selection of variables that includes all important ones.}
	\label{compare}
\end{figure}

The histograms depict the outcomes over all data sets in Figure \ref{compare}. In the top row we see the results if all variables were main effects only. The most successful criterion here is the one that uses leverage scores. This is to be expected based on what we observed in Figures \ref{part1} to \ref{part2b}. 

These stark differences in the quality of selected variables vanish for Scenarios 2 and 3 with the introduction of higher-order relationships between variables. That is, in the middle and bottom rows of Figure \ref{compare}, the histograms appear similar in shape. 

As interactions in the data move to a higher order, we see that for all criteria there is a shift in the histograms to the right. This means that the number of useful variables sampled from data simulated using Scenario 3, where we have up to fourth-order relationships, is higher than the number drawn from data simulated for Scenario 2. One does notice that, when relationships between variables exist, the subset of variables selected using cross-leverage scores finds at least 8-10 out of the 10 useful variables more often, on average, than all other approaches.

\subsection{Sampling success of leverage scores and cross-leverage scores combined}
\label{res:samp_success_combined}
Given that leverage scores clearly outperform cross-leverage scores when separating out main effects, we consider taking samples that use both approaches. In these samples, the first set of elements are a certain percentage of variables selected using cross-leverages, then a second set of elements are determined by repeating the exercise using leverage scores. For example, with a combination of 10\% CLS and 40\% LS, we sample 10\% of the variables using cross-leverage scores, and 40\% of the variables using leverage scores. It is possible, and often occurs, that there is an overlap in selected variables. 

In Table \ref{combo} we see the results from combined samples for 10\,000 simulated data sets with 60 observations and ten useful predictors, with up to fourth-order interactions between them, out of 1\,000 possible predictors. For each combination of LS and CLS, we measure the proportion of useful explanatory variables selected on average.
\begin{table}[ht]
	\centering
	\begin{tabular}{rl|cccccccccc}
		\hline
		&\multicolumn{11}{c}{ \% CLS}\\
		& &  0 &  10 & 20 & 30 & 40 & 50 & 60 & 70 & 80 & 90 \\ 
		\hline
		& 0  & \textbf{0.00} & 0.40 & \textbf{0.55} & 0.65 & \textbf{0.73} & 0.80 & \textbf{0.85} & 0.90 & \textbf{0.93} & 0.97 \\[2pt] 
		& 10 & 0.54 & \textbf{0.75} & 0.82 & \textbf{0.86} & 0.90 & \textbf{0.92} & 0.94 & \textbf{0.96} & 0.97 & \textbf{0.99} \\[2pt]
		& 20 & \textbf{0.62} & 0.80 & \textbf{0.85} & 0.89 & \textbf{0.91} & 0.93 & \textbf{0.95} & 0.97 & \textbf{0.98} & 0.99 \\[2pt] 
		& 30 & 0.69 & \textbf{0.83} & 0.88 & \textbf{0.91} & 0.93 & \textbf{0.95} & 0.96 & \textbf{0.97} & 0.98 & \textbf{0.99} \\[2pt]
		\% LS& 40 & \textbf{0.74} & 0.86 & \textbf{0.90} & 0.92 & \textbf{0.94} & 0.95 & \textbf{0.97} & 0.98 & \textbf{0.99} & 0.99 \\[2pt] 
		& 50 & 0.79 & \textbf{0.89} & 0.92 & \textbf{0.94} & 0.95 & \textbf{0.96} & 0.97 & \textbf{0.98} & 0.99 & \textbf{0.99} \\[2pt] 
		& 60 & \textbf{0.83} & 0.91 & \textbf{0.94} & 0.95 & \textbf{0.96} & 0.97 & \textbf{0.98} & 0.98 & \textbf{0.99} & 1.00 \\[2pt] 
		& 70 & 0.88 & \textbf{0.93} & 0.95 & \textbf{0.96} & 0.97 & \textbf{0.98} & 0.98 & \textbf{0.99} & 0.99 & \textbf{1.00} \\[2pt] 
		& 80 & \textbf{0.92} & 0.96 & \textbf{0.97} & 0.98 & \textbf{0.98} & 0.99 & \textbf{0.99} & 0.99 & \textbf{1.00} & 1.00 \\[2pt] 
		& 90 & 0.96 & \textbf{0.98} & 0.98 & \textbf{0.99} & 0.99 & \textbf{0.99} & 0.99 & \textbf{1.00} & 1.00 & \textbf{1.00} \\ 
		\hline
	\end{tabular}
	\vspace{0.1cm}
	\caption{Proportion of useful variables selected for various combinations of degree to which LS or CLS were used to sample. The bottom-left to top-right diagonals are highlighted to compare different mixture proportions where the total number of samples remains constant.}\label{combo}
\end{table}

In general, combining the scores is advantageous over the use of just one type of scores. For example, a sample of about 20\% of the original variables where half (10\%) are selected using LS and the other half (10\%) using CLS has a 75\% success rate, which is higher than 62\% and 55\% using LS and CLS only, respectively.

\subsection{Impact of sampling on subsequent statistical models}

In section \ref{res:samplingsuccess}, we compare the sampling success rates for the different variable selection methods. In this section, we focus on the next step: running a logic regression model on the original data set as well as on the subsets. To that end, we employ the R package logicFS \citep{Schwender2019} which is designed to identify influential higher-order interactions in addition to main effects.

We perform logic regression as implemented in the function \texttt{logicFS} on simulated data sets with $n=60$ and $p=1\,000$. Again, Scenarios 1, 2, and 3 are considered. In addition to variable selection via leverage scores, cross-leverage scores, correlations, and p-values, we add a combination of leverage and cross-leverage scores. Because the results in section \ref{res:samp_success_combined} indicate that selecting 10\% of the variables via leverage scores and the remainder via cross-leverage scores, we first choose $100$ variables with the lowest leverage scores. From the remaining 900 variables, a further 146 are then chosen on the basis of their cross-leverage scores.

We conduct logic regression on the original and all reduced data sets with the default values of the function \texttt{logicFS}, with the exception of the argument \texttt{nleaves=30}. For all logic regression models, we examine whether the main effect or the exact interaction has been found and if yes, what importance \texttt{logicFS} has assigned to it. We do not consider partial results like $X_{10} \land X_{25}$ in cases where $X_{10}$ is an important variable of its own. 

In our simulations, logic regression systematically finds more of the important variables on the reduced data sets compared to the large full data set, regardless of the method. The reduction of the search space seems to be beneficial compared to the risk of not finding all important variables in the reduction process.

Over the three scenarios, reducing the data sets employing (C)LS or a combination of both leads to \texttt{logicFS} finding more important main effects and interactions than employing correlations or p-values for the reduction.

Among the three strategies based on (C)LS, selection via cross-leverage scores leads to \texttt{logicFS} generally finding more interaction terms, but less main effects than selection via leverage scores. The combined strategy leads to results in between the two pure strategies with respect to both main effects and interactions. 

Surprisingly, selection via leverage scores seems to lead to better logic regression models even when the number of important variables found is the same. An illustrative example is Scenario 1, where both selection via leverage scores and via a combination of leverage and cross-leverage scores find all main effects in all cases. However, there is a difference in the logic regression models, which include the main effects in considerably more cases when the selection was done via leverage scores. Whether this happens by random chance or has an underlying cause, e.g. that the other selected variables are more distinctly uninfluential, is an open question.

\subsection{Analysis of HapMap data set}

As described in Section \ref{sec:HapMap}, the studied populations can be assigned to two different continents: Yorùbá/Africa (YRI) and central Europe (CEU). We first show a raster plot with the complete set of 7648 SNPs across the two groups of individuals included in our analysis. Raster plots are a graphical representation of a matrix containing SNP data. The rows correspond to the individuals and the columns to the SNPs. Three colors are chosen to represent values of 0, 1, and 2, respectively.
\begin{figure}[ht!] 
	\includegraphics[width=\textwidth]{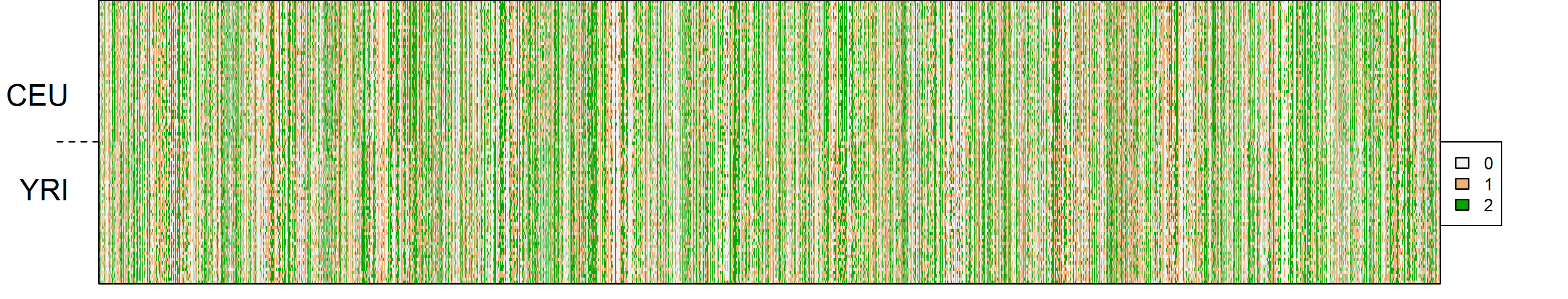}\\
	\caption[]{ Raster plots of all 120 subjects from two continental regions with respect to full set of 7648 SNPs (green/pink denotes homozygotic individuals and white denotes heterozygotic individuals). }
	\label{raster}
\end{figure}
\begin{figure}[ht!]
	\includegraphics[width=\textwidth]{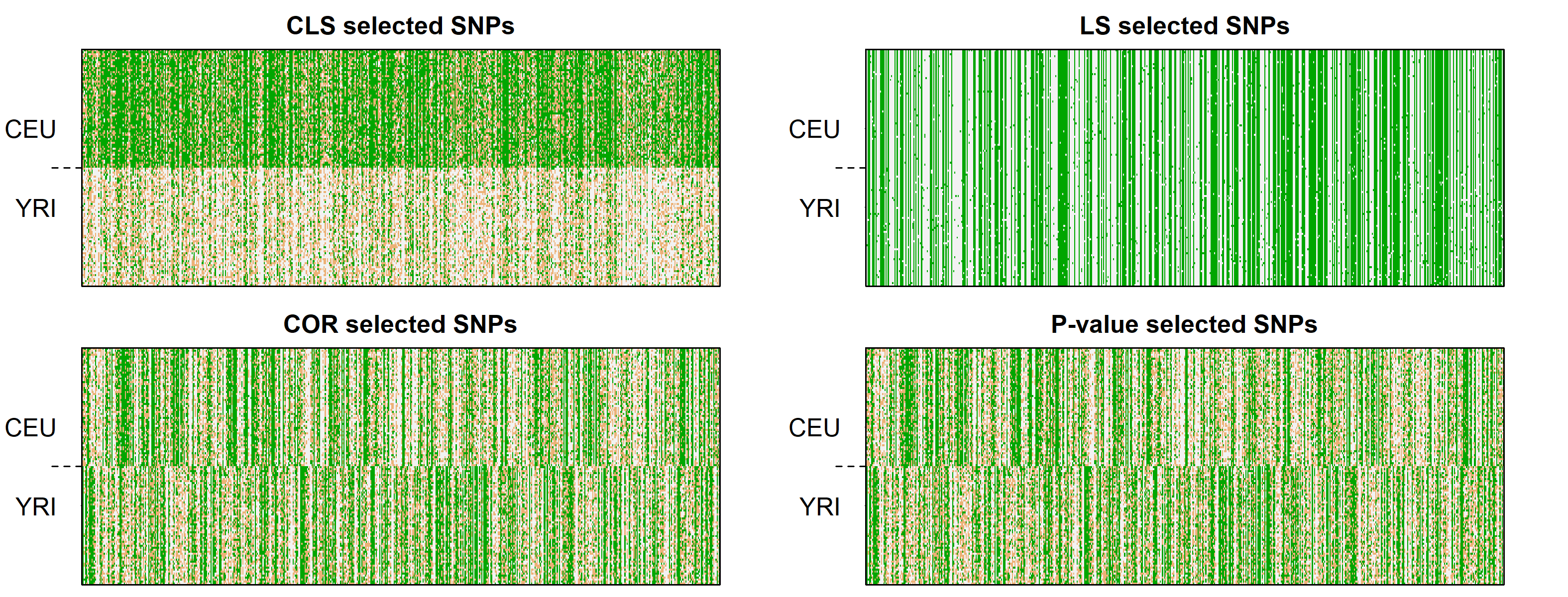}\\
	\caption[]{ Raster plots of all 120 subjects from two continental regions with respect to a subset of 575 out of 7648 SNPs (green/pink denotes homozygotic individuals and white denotes heterozygotic individuals), selected using CLS, LS, COR and p-values, respectively. }
	\label{fig6}
\end{figure}
A similar visualization is employed in \citet{Paschou2007}. In Figure \ref{raster} each of the first 60 rows index an individual in the CEU group and the final 60 rows correspond to an individuals in the YRI group. The columns corresponds to the each of the 7648 SNPs in the full data set.

Next, we produced a set of raster plots for the same 120 individuals, but for subsets of SNPs instead. The subsets are selected using each of our four criteria. We take a sample of size $\lceil 120\ln(120)\rceil = 575$ of the genetic markers (SNPs) in the data using cross-leverage scores, leverage scores, correlations and p-values, respectively. Figure \ref{fig6} shows the raster plots with the subsets of SNPs chosen for each of the criteria. 

We see that SNPs selected using cross-leverage scores display the most distinct patterns between the two groups, i.e. the two groups can easily be classified on the basis of the selected SNPs. The leverage scores show no blocking of the two groups. For selection according to correlations and p-values, there are hints of blocks corresponding to the two groups, but to a far lesser extent than what we see for the cross-leverage case.

\section{Discussion and further work}

Recent works in computer science have led to the development of highly efficient approaches to compute leverage scores and cross-leverage scores. We introduce the idea of using those measures -- especially CLS -- for variable selection and in particular for detecting variables that participate in \emph{interactions} effecting the response.

We derive an understanding of both types of scores when dealing with binary data. In summary, cross-leverages are distinct for useful variables when dealing with interactions and, to a lesser extent, main effects, while leverages are only distinct when contrasting main effects. 

When examining the effect of the number of variables, we found that both leverage and cross-leverage scores of useful variables do not become more distinct from irrelevant ones as $p$ is increased while the number of important variables remains fixed. However, both leverage and cross-leverage scores of useful variables become more distinct from irrelevant ones as $n$ increases, with cross-leverages improving more noticeably for higher $n$.

We show that the subset of variables captured by cross-leverage scores contain a larger number of useful variables on average. In general, both leverage and cross-leverage scores prove more useful than correlation scores and p-values. Since leverage scores outperform cross-leverages when dealing with main effects, we develop a variable selection algorithm that combines both measures with overall greater success than either measure on its own.

Table \ref{tab:overview} gives an overview of our key findings with regard to the variable selection process.

\begin{table}[bt]
	\centering
	\begin{tabular}{ll}
		\toprule
		Situation & Recommendation/Result\\
		\midrule
		main effect only & selection according to leverage scores\\
		interactions present & selection according to LS/CLS or combination\\
		growing $p$ & little influence on performance of selection\\
		growing $n$ & positive influence on performance of selection\\
		\bottomrule
	\end{tabular}
	\caption{Overview of findings with regard to the variable selection process.}
	\label{tab:overview}
\end{table}

We believe that cross-leverages are particularly useful when the main focus of interest is finding interaction terms. However, we also show here that including a pre-processing reduction step always presents an advantage when using logic regression, specifically the logicFS method. Specifically, the method runs faster, and has higher chance of finding meaningful interactions, since the reduction of dimensionality drastically reduces the exponentially large search space.

The structure in selected SNPs from the HapMap data were informative with respect to structural differences between the two populations considered. The two populations happen to be very distinct in nature, so there was no difference in terms of the ability to make predictions though.

In the presence of influential main effects or interactions we have always deterministically set $Y=1$. Future work could add noise, e.g. via probabilistic modeling with $P(Y=1| \text{interaction present})=0.95$. We expect that this would complicate finding important main effects and interactions. However, it is not obvious whether it would affect all the methods considered here in the same way.


Based on what we have presented here, we argue that cross-leverages may be considered a promising new tool for the selection of variables that can handle large wide data sets and uncover relationships between them that effect the response. The success of our selection approach is grounded in the fact that it takes relationships between explanatory variables into account when characterizing relevant variables. That is, interactions are treated as useful information instead of a nuisance. Our initial analysis on simulated and real-world data appears promising in terms of our method extracting meaningful subsets of SNPs for describing differences between genetic populations. This method will hopefully prove to be a noteworthy tool for researchers working with wide data such as genetic marker data.

\bibliographystyle{Chicago}
\bibliography{references}
\end{document}